\documentclass[10pt]{article}

\usepackage{graphicx,amssymb,amsmath,fullpage}
\usepackage[colorlinks,breaklinks,bookmarks,pdfauthor=
 MA Mosquera and WH Lizcano-Valbuena,pageanchor,citecolor=blue]{hyperref}

\title{Modeling of the anode side of a direct methanol fuel cell with analytical solutions}
\author{Mart\'in A. Mosquera$^a$, William H. Lizcano-Valbuena$^b$
\\\\$^a$School of Chemical Engineering, Universidad del Valle, Cali, Colombia\\$^b$Department of Chemistry, Universidad del Valle, Cali, Colombia}

\begin{document}
\maketitle

\begin{abstract}
In this work, analytical solutions were derived (for any methanol oxidation reaction order) for the profiles of methanol concentration and proton current density by assuming diffusion mass transport mechanism, Tafel kinetics, and fast proton transport in the anodic catalyst layer of a direct methanol fuel cell. An expression for the Thiele modulus that allows to express the anodic overpotential as a function of the cell current, and kinetic and mass transfer parameters was obtained. For high cell current densities, it was found that the Thiele modulus ($\phi^2$) varies quadratically with cell current density; yielding a simple correlation between anodic overpotential and cell current density. Analytical solutions were derived for the profiles of both local methanol concentration in the catalyst layer and local anodic current density in the catalyst layer. Under the assumptions of the model presented here, in general, the local methanol concentration in the catalyst layer cannot be expressed as an explicit function of the position in the layer. In spite of this, the equations presented here for the anodic overpotential allow the derivation of new semi-empirical equations. {\slshape Keywords:} Electrochemical kinetics; Catalyst layer; Mass transport; Methanol electro-oxidation; Proton transport.
\end{abstract}

\newcommand{\mr}[1]{\mathrm{#1}}
\newcommand{\mc}[1]{\mathcal{#1}}
\section{Introduction}
The direct methanol fuel cell (DMFC) has been proposed as an alternative clean and safe power source, due to the easy transportation and storing of liquid methanol and the corresponding low carbon dioxide emissions of the cell. However, the DMFC has performance inconveniences attributed to the methanol crossover through the solid polymer electrolyte (SPE) and kinetic limitations as the poisoning caused by species like carbon monoxide at the anode.\\
The simplest approach to the DMFC polarization curve is by using semi-empirical equations, and along with the nonlinear least-squares minimization technique, the polarization curve of a fuel cell (including a DMFC) can be fitted. One of these semi-empirical equations was first proposed by Srinivasan et. al \cite{ref15}:
\begin{equation}
E=E_{0}-b\ln\,I-R\,I\label{eq:siri}
\end{equation}
\begin{equation}
E_0=E_{\mr{r}}+b\ln\,I_0
\end{equation}
where $I$ is the cell current density and $E$ is the cell potential, and the other quantities are adjustable parameters. The equation of Srinivasan et al. \cite{ref15} is not suitable when mass transfer limitations occur. On the other hand, Kim et al. \cite{ref16} showed that the equation
\begin{equation}
E=E_0-b\ln I-RI-m\exp(nI)\label{eq:kim}
\end{equation}
accurately fits the polarization curve in the mass transfer region where the equation of Srinivasan et al. \cite{ref15} is not applicable; however, Eq. (\ref{eq:kim}) fails in the low cell current region. Considering this, Squadrito et al. \cite{ref20} proposed a semi-empirical equation in which the term $m\exp(nI)$ in Eq. (\ref{eq:kim}) was replaced by $aI^k\ln(1-\beta I)$; this modified equation works well over the whole cell current range. Additionally, semi-empirical models for low temperature fuel cells that are based on mass transport in the DMFC have been published. For example, by invoking an empirical mass transfer coefficient and Tafel kinetics, Argyropoulos et al. \cite{ref17} derived an equation with adjustable parameters. This equation accurately fits the polarization curve even in the low current region where other theoretical models fail; however, the semi-empirical equation derived by Argyropoulos et al. \cite{ref17} does not take into account the effects of polarization caused by the reduction of oxygen in the cathode. For this reason, Tu et al. \cite{ref21} proposed a semi-empirical equation that allows to include the effects in the polarization curve of both anodic and cathodic overpotentials, and the effect caused by the mixed potential in the cathode.\\
Equations derived from mathematical modeling are more rigorous than semi-empirical equations and yield more reliable results. One of the objectives of mathematical modeling is the prediction of fuel cell polarization curve, in which the effects of activation, and Ohmic and concentration overpotentials can be observed. Estimation of the effects of parameters is another objective of mathematical modeling. For this reason, it is necessary to propose models and perform simulations that make it possible to establish such effects on the fuel cell performance.\\
Baxter et al. \cite{ref9} considered the anode as a porous electrode covered by a selective SPE. By taking into account the transport of methanol, carbon dioxide, water and protons, and Butler-Volmer kinetics, a pseudo two-dimensional model was proposed, in which the catalyst layer consists of a liquid layer, a bond layer and a matrix layer. Baxter et al. \cite{ref9} considered the species exchange between the bond layer and the liquid layer; this mass transfer process is characterized by effective mass transfer coefficients. The results of Baxter et al. \cite{ref9} suggest that the anode thickness, the surface area, and the charge transfer coefficient are important for the determination of the anode behavior.\\
Scott et al. \cite{ref11} developed a model for the liquid-fed DMFC which considers the influence of the carbon dioxide, methanol and water transport at the fuel cell anode. The water transport is assumed to be caused by electro-osmotic drag and molecular diffusion. Those authors simulated the fuel cell polarization curve using a semi-empirical equation for the open circuit potential and a capillary pressure model. In this case, the spatial variation of methanol concentration in the catalyst layer was not considered.\\
Kulikovsky \cite{ref13} derived an analytical solution for proton transport across the cathode of a polymer electrolyte fuel cell (PEMFC), Tafel kinetics and no variation of the oxygen concentration throughout the cathodic catalyst layer were assumed; these solutions are appropriate for low cell current densities.\\
Wang et al. \cite{ref14} proposed a model for oxygen reduction at the electrocatalytic layer, in which the oxygen and proton transport occurs through spherical agglomerates, to study the case of fast proton transport, fast oxygen diffusion, and mixed control. For the case of low protonic conductivities, the results of Wang et al. \cite{ref14} suggest that the active part of the catalyst layer is located at the region close the membrane.\\
Many works have been published on modeling and simulation of fuel cells and are based on numerical methods \cite{ref9,ref14,ref10,ref12,ref52} to solve the transport equations, only few of those works have implemented analytical methods to simulate the fuel cell. However, the simulation of fuel cells does not dependent on the procedure of calculation, and analytical methods do not offer computational advantages over numerical methods. In spite of this, procedures based on analytical solutions contribute to the analysis and understanding of the processes that occur in the electrodes of fuel cells.\\
In this work, we have developed a one-dimensional analytical model of the anode side of a DMFC. Fick's law for methanol transport in the anode, and Tafel kinetics for methanol electro-oxidation, and fast proton transport were assumed. A non-linear differential diffusion-reaction equation for methanol transport, which was solved analytically, was obtained under these assumptions. The solution so obtained is an implicit function of the methanol concentration at the catalyst layer and it must be solved by an iterative procedure. 

\section{Theoretical model}
\label{theor}
In this model it is assumed that:
\begin{itemize}
\item[1.] The fuel cell temperature is constant.
\item[2.] The pressure difference between the both sides of the cell is negligible.
\item[3.] The membrane is fully hydrated.
\item[4.] The effects caused by carbon dioxide evaporation, methanol evaporation and water are negligible.
\item[5.] Carbon monoxide poisoning is negligible.
\item[6.] Steady state operation conditions.
\item[7.] Constant cathodic overpotential.
\item[8.] Methanol transport throughout the electrolyte phase in the catalyst layer is neglected.
\end{itemize}
\subsection{Methanol transport in CL}
\begin{figure}[H]
\centering
\includegraphics[width=0.7\textwidth]{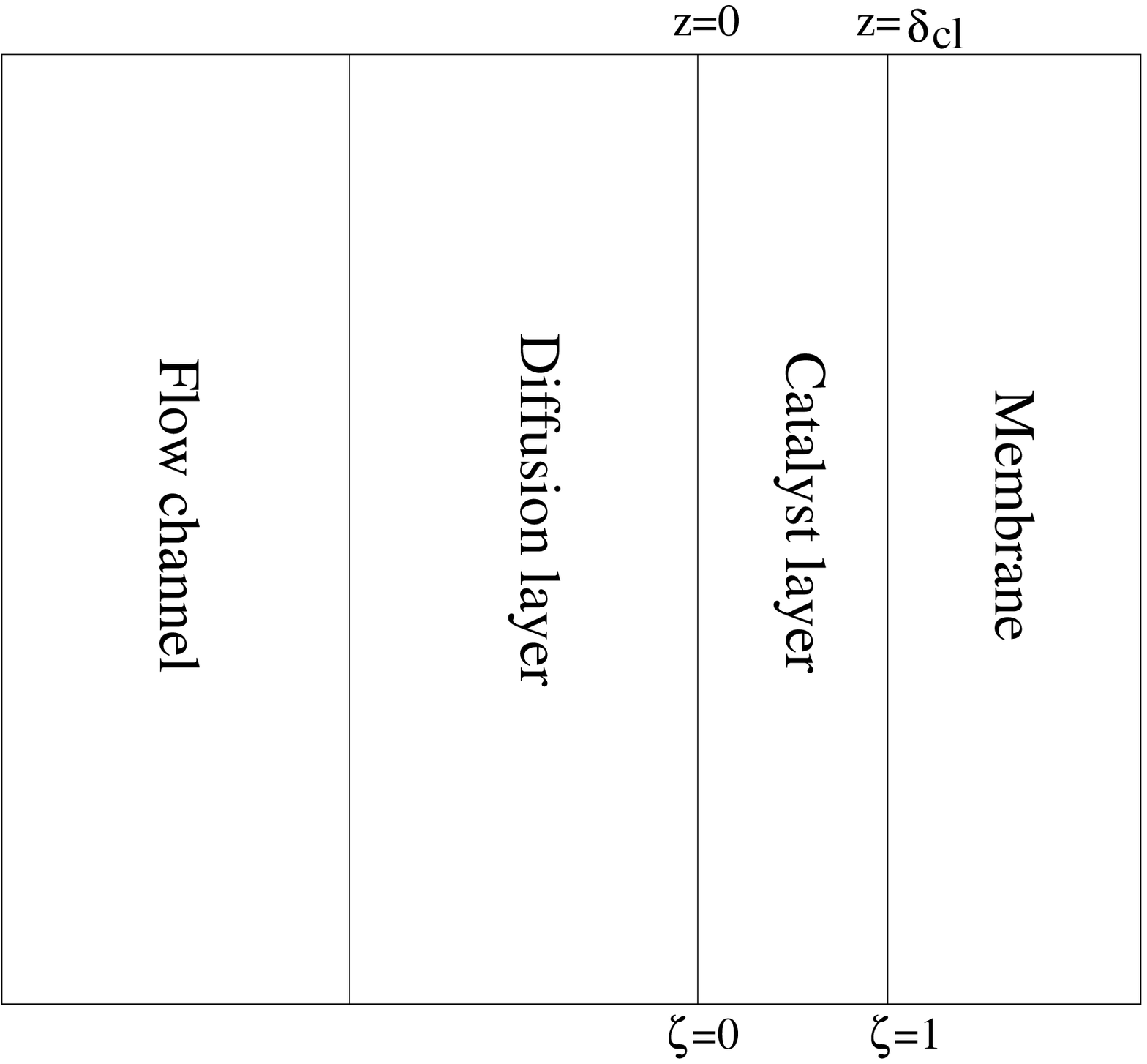}
\caption{}
\label{fig:fig111}
\end{figure}

Figure \ref{fig:fig111} shows a schematic representation of the anode side, this is divided into four zones: flow channel, diffusion layer (DL), catalyst layer (CL), and the membrane. The domain of the $z$ variable is the region $0<z<\delta_{\mr{cl}}$ in the CL. CL consists of a carbon supported solid catalyst. A finite external load induces three basic phenomena in the CL: (i) methanol transport, (ii) proton transport, and (iii) a finite overpotential on the anode. Methanol molecules in the DL are attracted towards the CL by electro-osmotic drag, and a concentration gradient caused by methanol electro-oxidation, once the molecules reach the CL, the electro-oxidation of methanol molecules decrease the methanol flux throughout the CL. In the membrane, methanol flux is considered constant. From the molecular diffusion flux definition (in steady state) it follows that 
\begin{equation}
-\mc{D}_{\mr{M}}^{\mr{cl,eff}}\frac{dC_{\mr{M}}^{\mr{cl}}}{dz}+v(z)C_{\mr{M}}^{\mr{cl}}(z)=N_{\mr{M}}^{\mr{cl}}(z)\label{eq:four}
\end{equation}
here, $C_{\mr{M}}^{\mr{cl}}$ is the local methanol concentration at the CL, $v$ is a convective velocity \cite{bern}, $\mc{D}_{\mr{M}}^{\mr{cl,eff}}$  is the methanol effective diffusion coefficient in the CL and it is given by the Bruggemann relation
\begin{equation}
\mc{D}_{\mr{M}}^{\mr{cl,eff}}=\epsilon_{\mr{cl}}^{3/2}\,\mc{D}_{\mr{M}}
\end{equation}

The methanol  flux is expressed as follows
\begin{equation}
N_{\mr{M}}^{\mr{cl}}(z)=\frac{I-j(z)}{n_{\mr{M}}\mc{F}}+N_{\mr{M}}^{\mr{m}}
\end{equation}
here, $I$ is the cell current density, $j$ is the local anodic current density, $N_{\mr{M}}^{\mr{cl}}$  is the local methanol flux at the CL, $N_{\mr{M}}^{\mr{m}}$ is the methanol crossover flux, and $n_{\mr{M}}$ is the number of electrons transferred per methanol molecule. 

\subsection{Analytical solutions for methanol transport in the CL}
Given that the convective velocity in the CL is small (no pressure gradients), the transport of methanol by convection mechanism in the CL can be neglected, and hence, Eq. (\ref{eq:four}) can be expressed as follows
\begin{equation}
\frac{dC_{\mr{M}}^{\mr{cl}}}{dz}\approx -\frac{N_{\mr{M}}^{\mr{cl}}(z)}{\mc{D}_{\mr{M}}^{\mr{cl,eff}}}
= \frac{j(z)}{n_{\mr{M}}\mc{F}\mc{D}_{\mr{M}}^{\mr{cl,eff}}}-\frac{N_{\mr{M}}^{\mr{d}}(z)}
{\mc{D}_{\mr{M}}^{\mr{cl,eff}}}\label{eq:eight}
\end{equation}
where $N_{\mr{M}}^{\mr{d}}$ is the methanol flux at the DL:
\begin{equation}
N_{\mr{M}}^{\mr{d}}=\frac{I}{n_{\mr{M}}\mc{F}}+N_{\mr{M}}^{\mr{m}}
\end{equation}
The volumetric anodic current ($dj/dz$) is given by 
\begin{equation}
\frac{dj}{dz}=Sj^{\mr{ref}}\Bigg(\frac{C_{\mr{M}}^{\mr{cl}}(z)}{C_{\mr{M}}^{\mr{ref}}}\Bigg)^{\gamma}
\exp\Big(\frac{\alpha \mc{F}}{RT}\eta\Big)\label{eq:five}
\end{equation}
where $S$ is the catalyst specific superficial area, $j^{\mr{ref}}$ is the reference current, $C_{\mr{M}}^{\mr{ref}}$ is the reference methanol concentration, $\alpha$ is the charge transfer coefficient, $\gamma$ is the methanol electro-oxidation reaction order, and $\eta$ is the anodic overpotential (assumed independent of $z$). Methanol flux is related with volumetric current as follows:
\begin{equation}
\frac{dN_{\mr{M}}^{\mr{cl}}}{dz}=-\frac{1}{n_{\mr{M}}\mc{F}}\frac{dj}{dz}
\end{equation}
substitution of this equation into Eq. (\ref{eq:five}) gives
\begin{equation}
\frac{d^2C_{\mr{M}}^{\mr{cl}}}{dz^2}-\frac{Sj^{ref}}{n_{\mr{M}}\mc{F}
\mc{D}_{\mr{M}}^{\mr{cl,eff}}}\exp\Big(\frac{\alpha \mc{F}}{RT}\eta\Big)\Bigg(\frac{C_{\mr{M}}^{\mr{cl}}(z)}{C_{\mr{M}}^{\mr{ref}}}\Bigg)^{\gamma}=0\label{eq:trans}
\end{equation}
Define the dimensionless variables
$$\zeta=\frac{z}{\delta_{\mr{cl}}}\quad \bar{C}(\zeta)=\frac{C_{\mr{M}}^{\mr{cl}}(\zeta)}{C_{\mr{M}}^{\mr{cl/dl}}}\quad J=\frac{j(\zeta)}{I}$$
$$\bar{\mc{I}}=\frac{I}{n_{\mr{M}}\mc{F}k_{\mr{cl}}C_{\mr{M}}^{\mr{cl/dl}}}\quad \bar{Q}(\zeta)=-\frac{d\bar{C}}{d\zeta}$$
The Thiele modulus is defined as
\begin{equation}
\phi^2=\frac{Sj^{\mr{ref}}\delta_{\mr{cl}}\exp\Big(\frac{\alpha \mc{F}}{RT}\eta\Big)}{n_{\mr{M}}\mc{F}k_{\mr{cl}}
C_{\mr{M}}^{\mr{cl/dl}}}\Bigg(\frac{C_{\mr{M}}^{\mr{cl/dl}}}{C_{\mr{M}}^{\mr{ref}}}\Bigg)^{\gamma}\label{eq:thiel}
\end{equation}
where $C_{\mr{M}}^{\mr{cl/dl}}$ is the methanol concentration at the CL/membrane interface, and
\begin{equation}
k_{\mr{cl}}=\frac{\mc{D}_{\mr{M}}^{\mr{cl,eff}}}{\delta_{\mr{cl}}}
\end{equation}
Define 
$$\bar{Q}_0=\bar{Q}(0)=\frac{N_{\mr{M}}^{\mr{d}}}{k_{\mr{cl}}C_{\mr{M}}^{\mr{cl/dl}}}\qquad \bar{Q}_1=\bar{Q}(1)=\frac{N_{\mr{M}}^{\mr{m}}}{k_{\mr{cl}}C_{\mr{M}}^{\mr{cl/dl}}} $$
and
$$\bar{C}_1=\bar{C}(1)=\frac{C_{\mr{M}}^{\mr{m}}}{C_{\mr{M}}^{\mr{cl/dl}}}$$
here $C_{\mr{M}}^{\mr{m}}$ is the methanol concentration at the CL/membrane interface. Note that
$$\bar{C}(0)=\frac{C_{\mr{M}}^{\mr{cl/dl}}}{C_{\mr{M}}^{\mr{cl/dl}}}=1$$
Using the previous dimensionless variables we can write Eq. (\ref{eq:five}) as
\begin{equation}
\frac{dJ}{d\zeta}=\frac{\phi^2}{\bar{\mc{I}}}\bar{C}^{\gamma}\label{eq:djdz}
\end{equation}
and Eq. (\ref{eq:trans}) as
\begin{equation}
\frac{d^2\bar{C}}{d\zeta^2}-\phi^2\bar{C}^{\gamma}=0\label{eq:six}
\end{equation}

Eq. (\ref{eq:six}) cannot be solved explicitly, but an implicit solution can be derived instead. Note that Eq. (\ref{eq:six}) can be expressed as 
\begin{equation}
\frac{d\bar{Q}^2}{d\zeta}-\frac{2\phi^2}{\gamma+1}\frac{d\bar{C}^{\gamma+1}}{d\zeta}=0\label{eq:app}
\end{equation}
Integration of Eq. (\ref{eq:app}) gives
\begin{equation}
\bar{Q}^2(\zeta)-\bar{Q}_1^2-\frac{2\phi^2}{\gamma+1}(\bar{C}^{\gamma+1}(\zeta)-\bar{C}_1^{\gamma+1})=0\label{eq:PP}
\end{equation}
This equation can be integrated as follows:
\begin{equation}
\begin{split}
\zeta&=\int^1_{\bar{C}}\frac{dX}{\displaystyle\sqrt{\bar{Q}_1^2+\frac{2\phi^2}{\gamma+1}
(X^{\gamma+1}-\bar{C}_1^{\gamma+1})}}\\
&=\frac{i}{\displaystyle\sqrt{\frac{2\phi^2}{\gamma+1}\bar{C}_1^{\gamma+1}-\bar{Q}_1^2}}\int^1_{\bar{C}}
{}_{\phantom{1}2}F_1\Bigg(\frac{1}{2},\beta;\beta
;\frac{X^{\gamma+1}}{\displaystyle\bar{C}_1^{\gamma+1}-\frac{\bar{Q}_1^2(\gamma+1)}{2\phi^2}}\Bigg)\,dX\\
&=\frac{-i}{\displaystyle\sqrt{\frac{2\phi^2}{\gamma+1}\bar{C}_1^{\gamma+1}-\bar{Q}_1^2}}
\Bigg[\bar{C}{}_{\phantom{1}2}F_1\Bigg(\frac{1}{2},\frac{1}{\gamma+1};1+\frac{1}{\gamma+1};\frac{\bar{C}^{\gamma+1}}
{\displaystyle \bar{C}_1^{\gamma+1}-\frac{(\gamma+1)\bar{Q}_1^2}{2\phi^2}}\Bigg)\\
&\qquad \qquad -{}_{\phantom{1}2}F_1\Bigg(\frac{1}{2},\frac{1}{\gamma+1};
1+\frac{1}{\gamma+1};\frac{1}
{\displaystyle \bar{C}_1^{\gamma+1}-\frac{(\gamma+1)\bar{Q}_1^2}{2\phi^2}}\Bigg)\Bigg]
\end{split}
\end{equation}
where ${}_{\phantom{1}2}F_1(a,b;c;x)$ is the hypergeometric function. An analytical continuation formula is required to compute the hypergeometric function \cite{whittaker,stegun}. Given that $\zeta$ is a real number, it follows that  
\begin{equation}
\begin{split}
\bar{C}\,\mr{Im}&\Bigg[{}_{\phantom{1}2}F_1\Bigg(\frac{1}{2},\frac{1}{\gamma+1};1+\frac{1}{\gamma+1};\frac{\bar{C}^{\gamma+1}}
{\displaystyle \bar{C}_1^{\gamma+1}-\frac{(\gamma+1)\bar{Q}_1^2}{2\phi^2}}\Bigg)\Bigg]\\-&\mr{Im}\Bigg[{}_{\phantom{1}2}F_1\Bigg(\frac{1}{2},\frac{1}{\gamma+1};
1+\frac{1}{\gamma+1};\frac{1}
{\displaystyle \bar{C}_1^{\gamma+1}-\frac{(\gamma+1)\bar{Q}_1^2}{2\phi^2}}\Bigg)\Bigg]=\zeta \sqrt{\frac{2\phi^2\bar{C}_1^{\gamma+1}}{\gamma+1}-\bar{Q}_1^2}\label{eq:hyper}
\end{split}
\end{equation} 
This solution holds for $2\phi^2\bar{C}_1^{\gamma+1}/({\gamma+1})-\bar{Q}_1^2>0$. If 
$$2\phi^2\bar{C}_1^{\gamma+1}/({\gamma+1})-\bar{Q}_1^2=0$$
it is obtained \cite{duong}
\begin{equation}
\bar{C}(\zeta)=\Big[1-\frac{(1-\gamma)}{2}\sqrt{\frac{2\phi^2}{\gamma+1}}\zeta\Big]^{\frac{2}{1-\gamma}}\label{eq:explicit}
\end{equation}
This is a special case where an explicit relation between $\bar{C}$ and $\zeta$ arises. For $\gamma=1$ we have
\begin{equation}
\bar{C}(\zeta)=\exp(-\phi\zeta)
\end{equation}
If 
$$2\phi^2\bar{C}_1^{\gamma+1}/({\gamma+1})-\bar{Q}_1^2<0$$
then we can integrate Eq. (\ref{eq:PP}) to give
\begin{equation}
\begin{split}
{}_{\phantom{1}2}F_1&\Bigg(\frac{1}{2},\frac{1}{\gamma+1};1+\frac{1}{\gamma+1};\frac{-1}
{\displaystyle\frac{(\gamma+1)\bar{Q}_1^2}{2\phi^2}-\displaystyle \bar{C}_1^{\gamma+1}}\Bigg)\\-\bar{C}\,&{}_{\phantom{1}2}F_1\Bigg(\frac{1}{2},\frac{1}{\gamma+1};
1+\frac{1}{\gamma+1};\frac{-\bar{C}^{\gamma+1}}
{\frac{\displaystyle(\gamma+1)\bar{Q}_1^2}{\displaystyle2\phi^2}- \displaystyle\bar{C}_1^{\gamma+1}}\Bigg)=\zeta \sqrt{\bar{Q}_1^2-\frac{2\phi^2\bar{C}_1^{\gamma+1}}{\gamma+1}}\label{eq:lowimpl}
\end{split}
\end{equation}
The profile of methanol concentration can be computed from the above equation by using an iterative procedure. In order to calculate the anodic polarization curve, it is necessary another relation between the local anodic current density and the local methanol concentration at the CL. From Eq. (\ref{eq:eight})
\begin{equation}
\frac{d\bar{C}}{d\zeta}=-\bar{Q}(\zeta)=\bar{\mc{I}}J(\zeta)-\bar{Q}_0\label{eq:fluxx}
\end{equation}
By invoking the chain rule:
\begin{equation}
\frac{dJ}{d\zeta}=\frac{dJ}{d\bar{C}}\frac{d\bar{C}}{d\zeta}
\end{equation}
we obtain
\begin{equation}
(\bar{\mc{I}}J-\bar{Q}_0)\frac{dJ}{d\bar{C}}=\frac{\phi^2}{\bar{\mc{I}}}\bar{C}^{\gamma}
\end{equation}
noting that $J(0)=0$ when $\bar{C}=1$, an integration of the above equation gives:
\begin{equation}
J^2-\frac{2\bar{Q}_0}{\bar{\mc{I}}}J+\frac{2\phi^2}{(\gamma+1)\bar{\mc{I}}}(1-\bar{C}^{\gamma+1})=0
\end{equation}
solving this second degree polynomial we obtain
\begin{equation}
J(\zeta)=\frac{\bar{Q}_0}{\bar{\mc{I}}}\Big[1-\sqrt{1-\frac{2\phi^2}{(\gamma+1)\bar{Q}_0^2}(1-\bar{C}^{\gamma+1}(\zeta))}\Big]\label{eq:curr}
\end{equation}

This equation relates the local anodic current density with the local methanol concentration at the CL. Also, Eq. (\ref{eq:curr}) allows to express local anodic current density as a function of the anodic overpotential, kinetic and mass transfer parameters. This equation is valid if the fast proton transport assumption is fulfilled. Substitution of this equation into Eq. (\ref{eq:fluxx}) gives
\begin{equation}
\frac{\bar{Q}(\zeta)}{\bar{Q}_0}=\sqrt{1-\frac{2\phi^2}{(\gamma+1)\bar{Q}_0^2}(1-\bar{C}^{\gamma+1}(\zeta))}\label{eq:Nflux}
\end{equation}
At the CL/DL interface Eq. (\ref{eq:PP}) becomes
\begin{equation}
\bar{Q}_0^2-\bar{Q}_1^2-\frac{2\phi^2}{\gamma+1}(1-\bar{C}_1^{\gamma+1})=0\label{eq:diff}
\end{equation}
It can be noted that Eq. (\ref{eq:eight}) can be written as
\begin{equation}
\bar{\mc{I}}=\bar{Q}_0-\bar{Q}_1\label{eq:IN0}
\end{equation}
Substitution of this equation into Eq. (\ref{eq:diff}) and rearranging we obtain
\begin{equation}
\phi^2=\frac{(\gamma+1)\bar{Q}_0^2}{2(1-\bar{C}_1^{\gamma+1})}\Big[1-\Big(1-\frac{\bar{\mc{I}}}{\bar{Q}_0}\Big)^2\Big]\label{eq:thiele1}
\end{equation}

Substitution of this equation into Eq. (\ref{eq:thiel}) gives the relation (using the original variables):
\begin{equation}
\eta=\frac{RT}{\alpha\mc{F}}\ln\Bigg\{\frac{1}{2}\frac{n_{\mr{M}}\mc{F}(N_{\mr{M}}^{\mr{d}})^2}{\displaystyle Sj^{\mr{ref}}\mc{D}_{\mr{M}}^{\mr{cl,eff}}C_{\mr{M}}^{\mr{ref}}}\frac{\gamma+1}{\displaystyle \Big[\Big(\frac{C_{\mr{M}}^{\mr{cl/dl}}}{C_{\mr{M}}^{\mr{ref}}}\Big)^{\gamma+1}-\Big(\frac{C_{\mr{M}}^{\mr{cl/m}}}{C_{\mr{M}}^{\mr{ref}}}\Big)^{\gamma+1}\Big]}
\Bigg[1-\Big(1-\frac{I}{n_{\mr{M}}\mc{F}N_{\mr{M}}^{\mr{d}}}\Big)^2\Bigg]\Bigg\}\label{eq:overpp}
\end{equation}

\subsection{Limiting cases}
\label{limit}

For high cell current density regime where $C_{\mr{M}}^{\mr{cl/m}}\rightarrow 0$, methanol flux crossover tends to zero and the following expression arises
\begin{equation}
\lim_{\bar{C}_1\rightarrow 0}\phi^2=\frac{\gamma+1}{2}\Big(
\frac{I}{n_{\mr{M}}\mc{F}k_{\mr{cl}}C_{\mr{M}}^{\mr{cl/dl}}}\Big)^2
\end{equation}
now it can be seen that $\phi^2\rightarrow \infty$ as $C_{\mr{M}}^{\mr{cl/dl}}\rightarrow 0$, and the anodic overpotential
\begin{equation}
\eta=\frac{RT}{\alpha\mc{F}}\ln\Bigg[\frac{\gamma+1}{2}\frac{n_{\mr{M}}\mc{F}k_{\mr{cl}}
C_{\mr{M}}^{\mr{cl/dl}}}{Sj^{\mr{ref}}\delta_{\mr{cl}}(C_{\mr{M}}^{\mr{cl/dl}}/C_{\mr{M}}^{\mr{ref}})^{\gamma}}\Big(
\frac{I}{n_{\mr{M}}\mc{F}k_{\mr{cl}}C_{\mr{M}}^{\mr{cl/dl}}}\Big)^2\Bigg]\label{eq:asin}
\end{equation}
tends to infinity. This approximation avoids the calculation of the hypergeometric function and the iterative calculations for high fuel cell current densities. If $\bar{C}_1\rightarrow 0$, Eq. (\ref{eq:PP}) can be integrated to give Eq. (\ref{eq:explicit}). Therefore, Eq. (\ref{eq:explicit}) applies for high current densities and for the special case where $2\phi^2\bar{C}_1^{\gamma+1}/({\gamma+1})-\bar{Q}_1^2=0$.

\subsection{A simple model for methanol crossover}
\label{simplem}
In order to simulate dimensionless quantities, the methanol crossover by electro-osmotic drag is neglected. Thus, 
\begin{equation}
\bar{Q}_1=\frac{k_{\mr{m}}}{k_{\mr{cl}}}\bar{C}_1=r\bar{C}_1
\end{equation}
here
\begin{equation}
k_{\mr{m}}=\frac{\mc{D}^{\mr{m}}_{\mr{M}}}{\delta_{\mr{m}}}
\end{equation}
where $\mc{D}^{\mr{m}}_{\mr{M}}$ is the methanol diffusion coefficient in the membrane and $\delta_{\mr{m}}$ is the diffusion layer thickness. This expression is used only for computational purpose; the ratio $r$ was taken as 0.08 \cite{ref52}. Electro-osmosis and pressure gradients must be included for realistic simulations.

\section{Discussion}

\begin{figure}[H]
\centering
\includegraphics[width=1.2\textwidth]{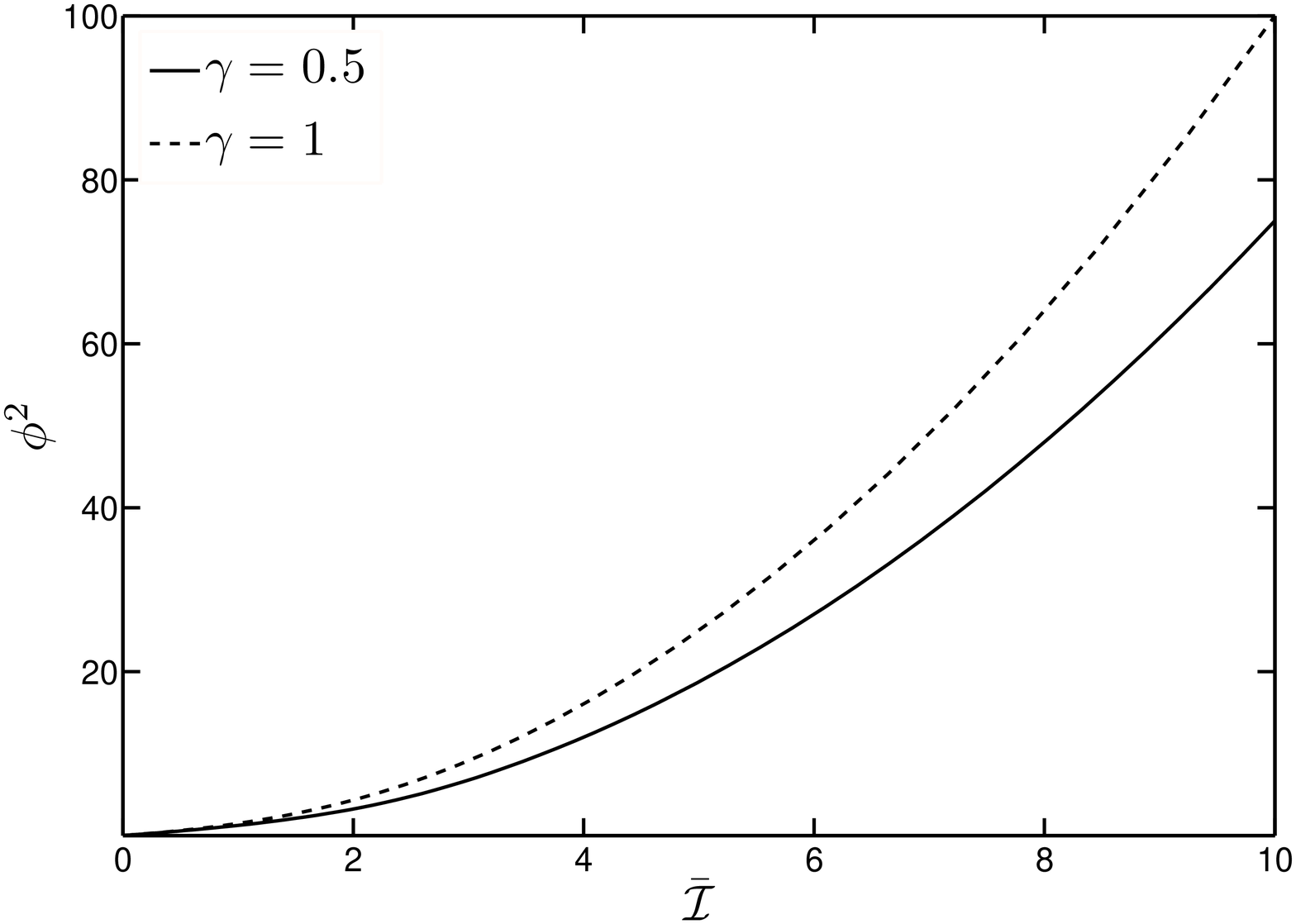}
\caption{}
\label{fig:fig1}
\end{figure}

\begin{figure}[H]
\centering
\includegraphics[width=1.15\textwidth]{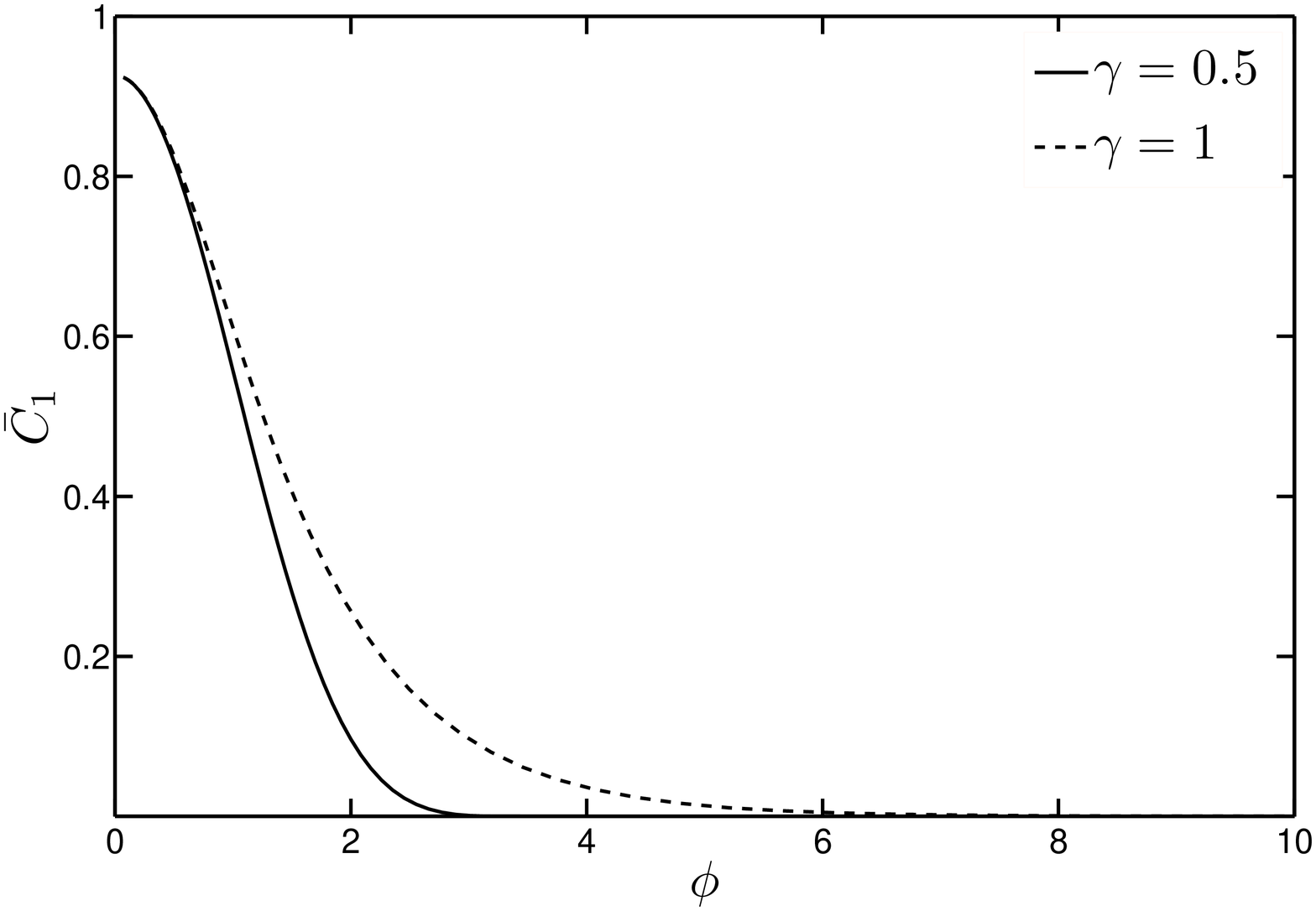}
\caption{}
\label{fig:fig2}
\end{figure}

\begin{figure}[H]
\centering
\includegraphics[width=1.15\textwidth]{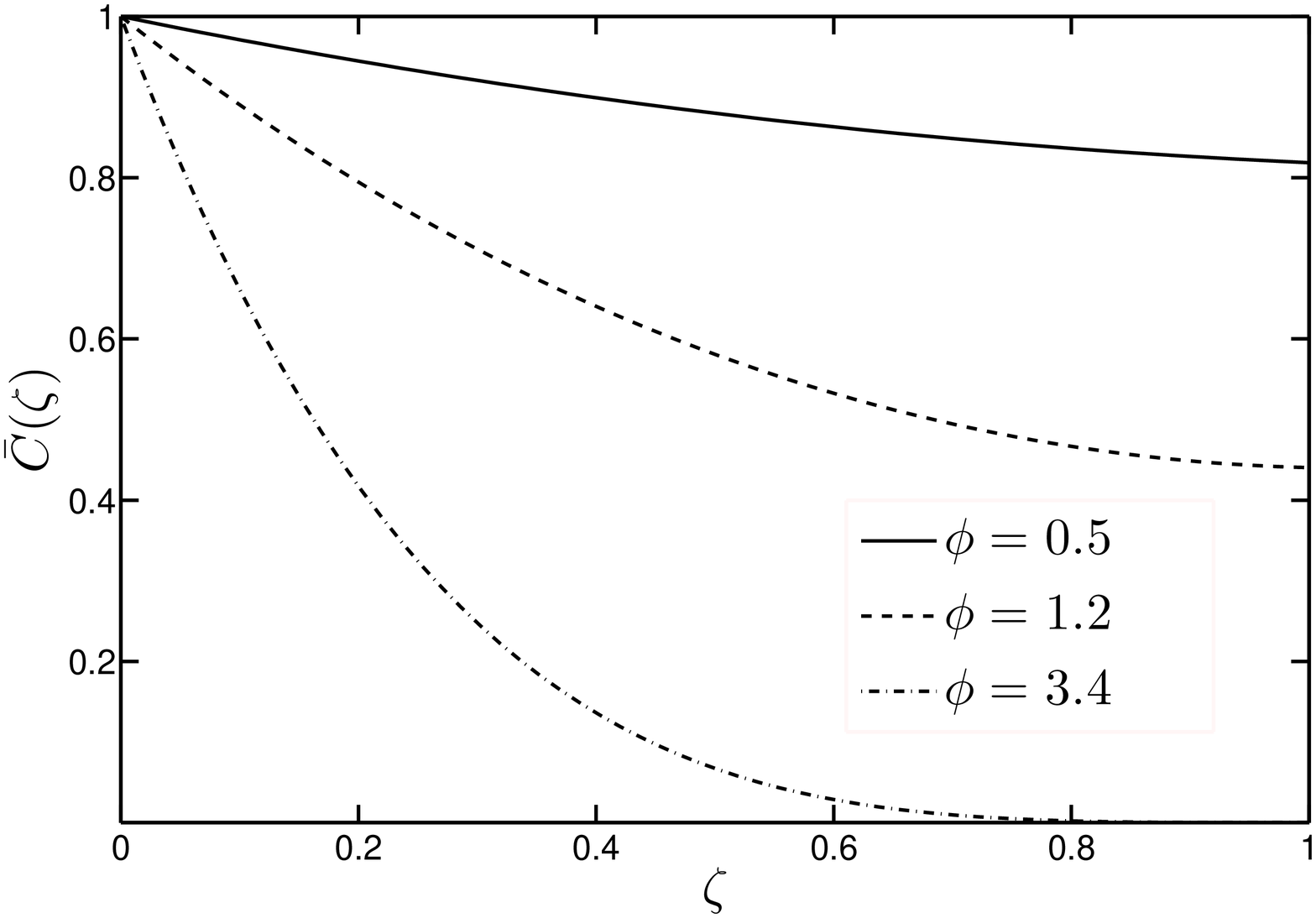}
\caption{}
\label{fig:fig3}
\end{figure}

\begin{figure}[H]
\centering
\includegraphics[width=1.15\textwidth]{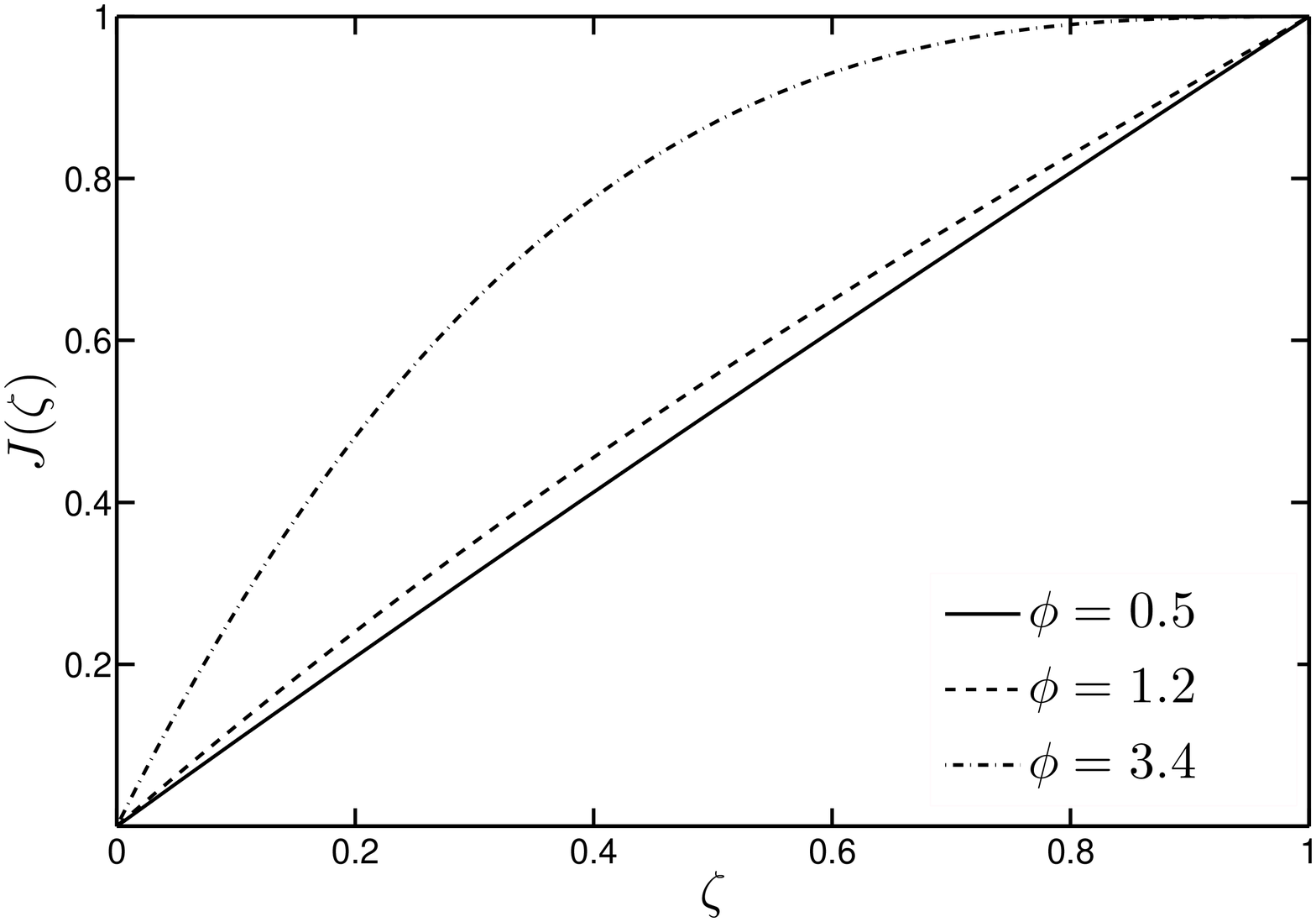}
\caption{}
\label{fig:fig4}
\end{figure}

\begin{figure}[H]
\centering
\includegraphics[width=1.15\textwidth]{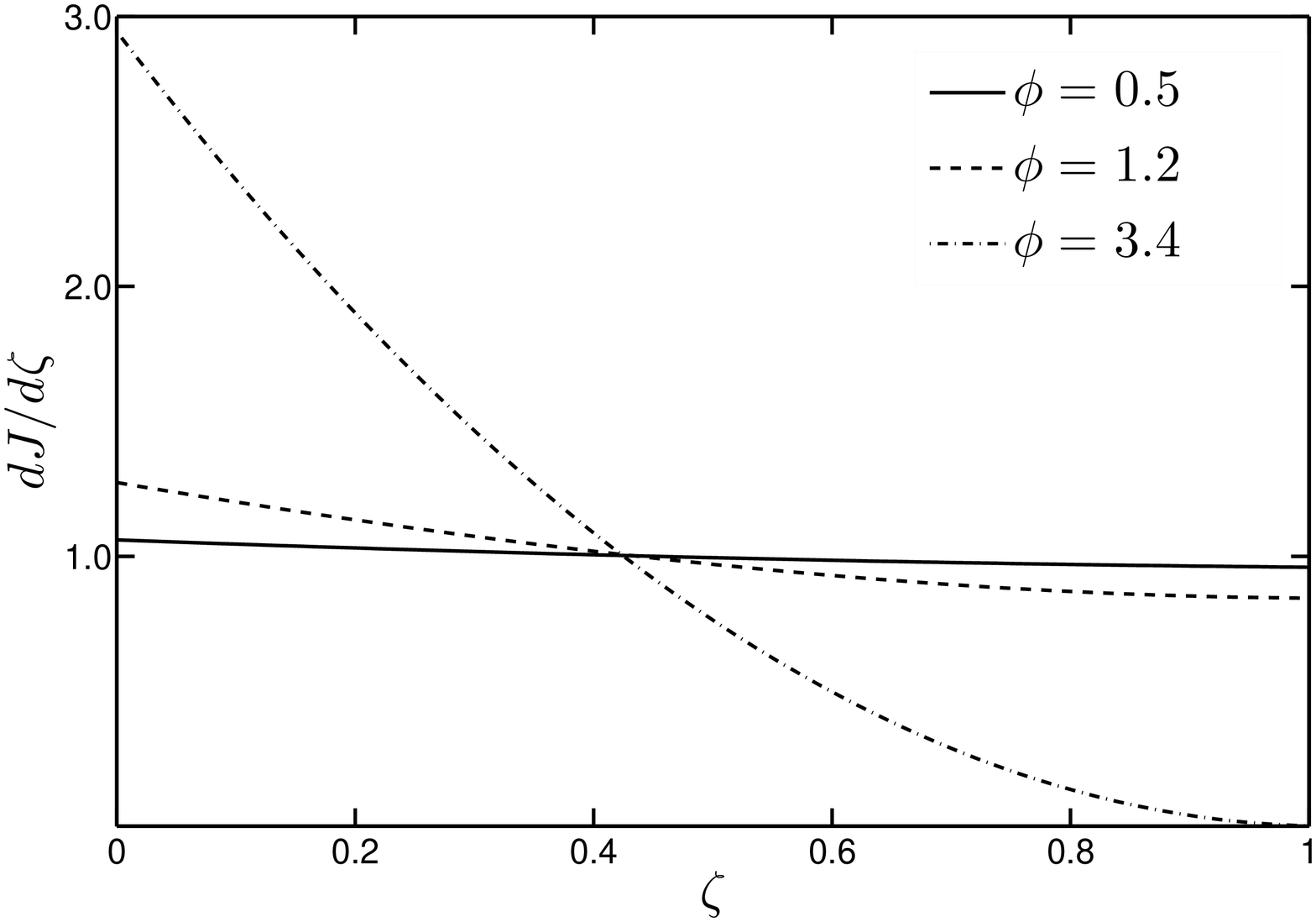}
\caption{}
\label{fig:fig5}
\end{figure}

In order to illustrate the analytic method presented in this work, curves of the Thiele modulus ($\phi^2$) for $\gamma=0.5$ and $\gamma=1$ are shown in Fig. \ref{fig:fig1}. As expected, the Thiele modulus tends to increase quadratically in the mass transfer region (i.e., $\phi\sim 4$) where the methanol crossover flux tends to zero. In this region, the approximation (Eq. (\ref{eq:asin})) shown in Sec. \ref{limit} is applicable in this model, and easily allows to estimate the anodic polarization curve for high fuel cell current densities. For low and moderate cell currents (i.e., $0<\phi<2$), Eq. (\ref{eq:overpp}) must be used. This Eq. gives the anodic overpotential as a function of methanol concentration at the CL/membrane interface and cell current density. Due to the fact that the hypergeometric function in Eqs. (\ref{eq:hyper}) and (\ref{eq:lowimpl}) cannot be inverted (except for some special cases), it is not possible  to express the methanol concentration at the CL in terms of position. This implies that the methanol concentration at the CL/membrane interface cannot be expressed as an explicit function of the fuel cell current. Now, given that the anodic overpotential in Eq. (\ref{eq:overpp}) depends on the methanol concentration at the CL/membrane interface, it cannot be expressed as an exact explicit function of the fuel cell current uniquely. An alternative is to neglect the methanol concentration at the CL/membrane, suggesting that the Eq. (\ref{eq:asin}) could be used to fit experimental data. However, because Eqs. (\ref{eq:asin}) and (\ref{eq:overpp}) do not correspond one to another, neglecting the methanol concentration at the CL/membrane interface could limit the applicability of Eq. (\ref{eq:asin}) for fitting data.\\
Fig. \ref{fig:fig2} shows plots of dimensionless methanol concentration at the CL/membrane interface vs. Thiele modulus ($\phi$). It can be noticed that the decrease of dimensionless methanol concentration at the CL/membrane interface is more pronounced as the reaction order is reduced, giving a reduction of methanol crossover. Fig. \ref{fig:fig2} can be interpreted as the diffusive part of the dimensionless methanol crossover flux. For a more detailed analysis of methanol crossover flux, electro-osmosis and pressure gradients must be added to the diffusive contribution of the methanol crossover flux   \cite{ref12,Liuu}.\\
The dimensionless profiles of local anodic current density for various Thiele modulus are shown in Fig. \ref{fig:fig3}, for moderate current densities, the profile shows a linear behavior where the methanol concentration in the CL is appreciable and nearly constant across the CL. This suggests that the proton generation is homogeneous in the CL. The local anodic current density becomes non-linear in the concentration overpotential region. In this case, the methanol concentration is negligible in the vicinity of the membrane, and the proton flux is generated mainly in the vicinity of the DL, and it becomes constant near the CL/membrane interface. Wang et al. \cite{ref14} reported similar results for the dimensionless profiles of current density for the case of oxygen transport in a cathodic catalyst layer by assuming oxygen transport through spherical agglomerates and fast proton transport.\\
The profiles of local methanol concentration at the CL are shown in Fig. \ref{fig:fig4}, in the activation region there is a little concentration difference between $\zeta=0$ and $\zeta=1$. In accordance with the results shown in Fig. \ref{fig:fig3}, Fig. \ref{fig:fig4} show that in the mass transport limitation region, the methanol concentration at the CL/membrane interface and methanol crossover (Fig. \ref{fig:fig2}) becomes negligible \cite{ref12}. In this case Eq. (\ref{eq:explicit}) is suitable to calculate local methanol concentration at the CL profile.\\
Finally, the Fig. \ref{fig:fig5} shows the profiles of dimensionless electrochemical reaction rate in the CL, these profiles can be related to the CL activity, for low current densities the activity is uniform across the CL, for high current densities, most of the activity of the CL is located near the DL/CL interface, and it controls the fuel cell performance, this confirms the characteristics of the Figs. \ref{fig:fig3} and \ref{fig:fig4}. This analysis of the electrochemical reaction rate profiles is applicable for the reduction of oxygen at the cathode of a PEMFC \cite{bern}.\\
Eq. (\ref{eq:six}) relates the methanol reaction and diffusion in the CL of the DMFC. The analytical solutions of this equation (Eqs. (\ref{eq:hyper}),(\ref{eq:explicit}) and (\ref{eq:lowimpl})) are in general implicit functions (except Eq. (\ref{eq:explicit})) of the local methanol concentration, the analysis of these equations is not easy, and it requires an iterative procedure to calculate the methanol profile in the CL.  Once calculated the dimensionless methanol concentration profile, the local anodic current density can be computed by using Eq. (\ref{eq:curr}). This Eq. gives a simple relationship between current density and the methanol concentration. It can be noted that Eq. (\ref{eq:curr}) apparently shows that the profile of local anodic current density is non-linear. However, a Taylor expansion of Eq. (\ref{eq:curr}) can show that the local anodic current density is a linear function of $\zeta$ for low current densities. Therefore, in the special case of low cell current densities, the profiles of local methanol concentration at the CL and local anodic current density can be expressed as linear functions of $\zeta$.\\
Fast proton transport ($d\eta/dz=0$) was the key assumption to derive Eq. (\ref{eq:hyper}), if the electrical resistance of the CL is negligible, then the local anodic overpotential is nearly constant; this occurs when the CL is very thin and its pores are well covered by the SPE. Consequently, the analytical solutions presented in this work cannot be applied if the fast proton transport assumption is not fulfilled. For the case of oxygen reduction, the simulation presented by Bernardi and Vebrugge \cite{bern} shows a nearly constant overpotential profile in the cathodic catalyst layer, and Gurau et al. \cite{gurau} take advantage of this fact to derive analytical solutions in order to simulate a porous electrode with $\gamma=1$.\\
The analytic model for the anodic catalyst layer presented here is subjected to the limitations of other one-dimensional models (see \cite{ref10,garcia,chenyeh}) that do not consider two-phase transport. For example, Jeng and Chen \cite{ref10} modelled the anodic catalyst layer with Tafel Kinetics and two-phase transport was neglected. In their model, like ours, one-dimensional transport of methanol in the anodic catalyst layer was considered. They showed that their model accurately fits the anodic polarization curve for 2 M of methanol feed concentration. Nevertheless, further validation of their model was not reported.\\
Another model that considers spatial variations of methanol concentration in the anodic catalyst layer of a DMFC was reported by Garc\'ia et. al \cite{garcia}. But, in comparison with the model of Jeng and Chen \cite{ref10}, Garc\'ia et. al \cite{garcia} modelled the anodic catalyst layer of a DMFC using the kinetic expression of Meyers and Newmann \cite{meyers}, and it was assumed that the local anodic current density and the local anodic overpotential were independent of position. The simulation of the polarization curve reported by Garc\'ia et. al \cite{garcia} shows good agreement with the experimental data only for 0.05, 0.1, 0.2 and 0.5 M of methanol feed. They found that for methanol feed concentrations greater than 0.5 M it was necessary another set of fitted kinetic parameters to model the polarization curve, and they suggest that this was caused by neglecting two-phase transport. In fact, it has been shown that models without gas transport do not agree well with experimental data for high fuel cell current densities \cite{ref12,Liuu}. Therefore, in order to perform more accurate simulations, important phenomena like two-phase transport \cite{ref12,Liuu}, transport of species through the electrolyte \cite{ref9,sund}, energy transport \cite{sund}, mass transport along the flow channels \cite{Birgen}, and non-Tafel kinetics must be included in a rigorous DMFC model. Although these phenomena were not considered in this work, the equations derived here can be used for rapid estimations, approximate quantitative analysis of the performance of the DMFC, and derivation of semi-empirical equations.

\section{Conclusions}
In this work, analytical solutions for the anode side of a DMFC performance simulation were presented. These solutions are applicable to the case of  negligible CL proton resistance and one-dimensional transport. However, the equations obtained for the methanol concentration and local anodic current density profiles are not explicit functions of the position in the CL. This implies that an iterative procedure must be used to obtain these profiles.\\
A set of analytical equations that relate the local methanol concentration at the CL with dimensionless position was found. In general, these equations do not allow to find an expression of the local methanol concentration at the CL as an explicit function of dimensionless position. Because of this, analytical expressions for local anodic current density and methanol flux at the CL as function of local methanol concentration at the CL were derived. Therefore, if the profile of local methanol concentration at the CL is known, then the local anodic current density and methanol flux at the CL profiles can be easily calculated by means of Eqs. (\ref{eq:curr}) and (\ref{eq:Nflux}) respectively.\\ 
Thiele modulus was found to be a function only of mass transport quantities such as the methanol flux in the DL and the methanol concentration in the CL/membrane interface. This dependency is caused by the assumption of fast proton transport, if this assumption is not fulfilled, the expressions derived for the Thiele modulus are no longer applicable, and there is no analytical method available to simulate the cell performance; hence, a procedure based on numerical methods would be required instead. \\
The expression derived in this work for the Thiele modulus allows the calculation of the anodic overpotential, and the profiles of methanol concentration and current density in the CL. Also, by means of the general expression for the Thiele modulus derived (Eq. (\ref{eq:thiele1})), the case of high fuel cell current densities was studied. A simple expression that relates the anodic overpotential with cell current density was obtained (\ref{eq:asin}). This expression could be used to fit experimental polarization curves, but it might not work for low cell current densities.\\

\pagebreak

\newpage

\newpage
\appendix

\section{Notation}

\begin{tabular}{ll} 
$\bar{C}$ & dimensionless methanol concentration as function of dimensionless position in CL\\
$\bar{C}_0$ &dimensionless methanol concentration at the CL/DL interface\\
$\bar{C}_1$ & dimensionless methanol concentration at the CL/membrane interface\\
$C_{\mr{M}}^{\mr{cl}}$ & methanol concentration in CL as function of position $(\mr{mol}/\mr{cm^3})$ \\ 
$C_{\mr{M}}^{\mr{cl/dl}}$ & methanol concentration at the CL/DL interface $(\mr{mol}/\mr{cm^3})$ \\ 
$C^{\mr{cl/m}}_{\mr{M}}$ & methanol concentration at the CL/membrane interface $(\mr{mol}/\mr{cm^3})$ \\ 
$C_{\mr{M}}^{\mr{ref}}$  & reference methanol concentration $(\mr{mol}/\mr{cm^3})$ \\ 
$\mc{D}_{\mr{M}}$ & diffusion coefficient of methanol in water $(\mr{cm^2}/\mr{s})$ \\ 
$\mc{D}_{\mr{M}}^{\mr{cl,eff}}$ & effective diffusion coefficient of methanol in CL $(\mr{cm^2}/\mr{s})$ \\  
$\mc{D}_{\mr{M}}^{\mr{m}}$ & diffusion coefficient of methanol in the membrane $(\mr{cm^2}/\mr{s})$ \\ 
$\mc{F}$ & Faraday constant $(\mr{C}/\mr{mol})$ \\ 
$I$ & cell current $(\mr{A}/\mr{cm^2})$   \\ 
$\bar{\mc{I}}$ & dimensionless cell current \\
$j$ & local anodic current density  $(\mr{A}/\mr{cm^2})$ \\ 
$j^\mr{{ref}}$ & reference exchange current density  $(\mr{A}/\mr{cm^2})$ \\ 
$J$ & dimensionless local current density \\
$k_{\mr{cl}}$ & mass transfer coefficient of methanol in the CL  $(\mr{cm}/\mr{s})$ \\ 
$k_{\mr{m}}$ & mass transfer coefficient of methanol in the membrane  $(\mr{cm}/\mr{s})$ \\ 
$n_{\mr{M}}$ & number of electrons transferred per methanol molecule  \\
$\bar{Q}$ & dimensionless methanol flux as function of dimensionless position\\  
$\bar{Q}_0$ & dimensionless methanol concentration at the CL/DL interface\\
$\bar{Q}_1$ & dimensionless methanol concentration at the CL/membrane interface\\
$N_{\mr{M}}^{\mr{cl}}$ & local methanol flux in CL $(\mr{mol}/\mr{cm^2\,s})$ \\ 
$N_{\mr{M}}^{\mr{d}}$ & methanol flux in DL $(\mr{mol}/\mr{cm^2\,s})$ \\ 
$N_{\mr{M}}^{\mr{m}}$ & methanol flux in M $(\mr{mol}/\mr{cm^2\,s})$ \\ 
$R$ & ideal gas constant $(\mr{J}/\mr{mol\,K})$ \\ 
$S$ & CL specific area $(\mr{cm^2}/\mr{cm^3})$ \\ 
$T$ & cell temperature $(\mr{K})$ \\ 
$v$ & convective velocity\\ 
$z$ & position in CL\\
$\alpha$ & charge transfer coefficient \\ 
$\gamma$ & reaction order of methanol electro-oxidation \\ 
$\delta_{cl}$ & thickness of CL $(\mr{cm})$ \\
$\delta_{m}$ & thickness of m $(\mr{cm})$ \\  
$\epsilon_{\mr{cl}}$ & CL void fraction \\ 
$\eta$ & anodic overpotential $(\mr{V})$ \\ 
$\phi^2$ & Thiele modulus \\ 
$\zeta$ & dimensionless position in CL.
\end{tabular}
\newpage
{\bfseries Figures:}\\
Figure 1. Schematic representation of the anode side of a DMFC.\\
Figure 2. Plots of Thiele modulus vs. dimensionless current for $\gamma=0.5$ and $\gamma=1$.\\
Figure 3. Plots of dimensionless methanol concentration at the CL/membrane vs Thiele modulus for $\gamma=0.5$ and $\gamma=1$.\\
Figure 4. Dimensionless methanol concentration profiles for various Thiele modulus; $\gamma=0.5$.\\
Figure 5. Dimensionless local anodic current density profiles for various Thiele modulus; $\gamma=0.5$.\\
Figure 6. Electrochemical reaction rate profiles for various Thiele modulus; $\gamma=0.5$.
\end{document}